\documentclass[rmp,aps,groupedaddress]{revtex4-2}

\usepackage{graphicx}   
\usepackage{amsmath}    
\usepackage{amssymb}    
\usepackage{bm}         

\usepackage[T1]{fontenc}
\usepackage{comment}

\usepackage{amsmath}
\usepackage{amssymb}
\usepackage{amsthm}
\usepackage{mathtools}

\newtheorem{corollary}{Corollary}
\newtheorem{definition}{Definition}
\newtheorem{lemma}{Lemma}
\newtheorem{principle}{Principle}
\newtheorem{proposition}{Proposition}
\newtheorem{theorem}{Theorem}

\newcommand{\QSTAB}{\operatorname{QSTAB}}

\newcommand{\1}{\mathbf{1}}

\newcommand{\Ex}{Ex}

\newcommand{\sfm}{\mathsf m}
\newcommand{\sfn}{\mathsf n}
\newcommand{\sfS}{\mathsf S}

\newcommand{\PP}{\mathcal P}
\newcommand{\OO}{\mathcal O}

\setcitestyle{super}  

\begin{document}



\title{Plenitude of ideal correlations explains quantum theory}


\author{Ad\'an Cabello}
\affiliation{Departamento de F{\'\i}sica Aplicada II, Universidad de Sevilla, E-41012 Sevilla, Spain}
\affiliation{Instituto Carlos I de F{\'\i}sica Te\'orica y Computacional, Universidad de Sevilla, E-41012 Sevilla, Spain}

\date{\today}


\begin{abstract}
We show that the Hilbert-space formalism and the Born rule of quantum theory are singled out by the following principle: every correlation that is possible in a theory in which every observable admits an ideal measurement is physically realizable. An ideal measurement is one that yields the same outcome when repeated and does not disturb any compatible observable.
\end{abstract}

\maketitle

\tableofcontents


\section{Motivation and structure}


The greatest quantum mystery is quantum theory itself: Where does it come from? What is the deeper foundation underneath it? \cite{Bernstein1989Wheeler} To shed light on these questions, we consider the following problem:

\begin{quote}
\textbf{Problem 1.} What is the physical principle that singles out quantum correlations in every Bell scenario?
\end{quote}

So far, no single, compelling physical principle has been found that reproduces the set of quantum correlations across all Bell scenarios. While several principles can bound the quantum set in some scenarios, all of them fail to exactly select the quantum set for all scenarios\cite{Popescu:1994FPH,vanDam:2000,vanDam:2013NC,Brassard:2006PRL,Pawlowski:2009NAT,Navascues:2010PRSA,Oppenheim:2010SCI,Fritz2013,Sun:2019,Miklin:2021PRL,Gachechiladze:2022,Jain:2024PRL,Gachechiladze:2026,Pompili:2026,Ramanathan:2026}.

We argue that Problem~1 cannot be solved by treating quantum correlations scenario by scenario. Such an approach is unnatural from the perspective of fundamental physical principles. Any Bell scenario can be extended in infinitely many ways---for example, by adding parties, measurement settings, or outcomes; by permitting sequential measurements; or by running a different experiment in parallel with it. A fundamental principle that accounts for the quantum correlations in one Bell scenario should also account for them in every extension of that scenario; in particular, the correlations of the smaller scenario should be recoverable as marginals of correlations in the larger one. The correlation sets of all scenarios are thus related by projection and embedding maps. Yet the rules governing these connections are not encoded in the definition of a scenario---its number of parties, measurement settings, and outcomes. A fundamental principle should produce the full compositional structure relating all scenarios, rather than the correlation set of each isolated scenario. 

The question is how to capture these connections between correlation sets across scenarios while preserving the fundamental appeal of Problem~1.

Let us recall why we regard Problem~1 as so important. Each Bell scenario has its own distinct set of quantum correlations. In every Bell scenario, the set of quantum correlations has a curved boundary with rich structure\cite{Goh:2018PRA,Le:2023Q,Barizien:2025NP}. On the one hand, this geometric complexity makes the quantum correlation sets difficult to characterize by a finite number of linear inequalities; on the other hand, the collection of all such sets provides a detailed and distinctive signature of quantum theory. It may therefore be the case that the principle singling out quantum correlations across all Bell scenarios is the principle underlying quantum theory.

How, then, should we proceed? We propose an indirect approach to Problem~1. Rather than attacking it scenario by scenario, we consider the following broader problem:

\begin{quote}
\textbf{Problem 2.} What is the physical principle that singles out quantum correlations in every exclusivity graph?
\end{quote}

For the definition of exclusivity graph, see Sec.~\ref{sec:graphs}. This reformulation of Problem~1 is natural because every Bell scenario is associated with a particular exclusivity graph, and exclusivity graphs provide a unified language in which the projection and embedding relations between scenarios can be expressed. A principle that solves Problem~2 would therefore capture connections that a scenario-by-scenario approach misses.

In addition, the following observations motivate our interest in Problem~2:
\begin{itemize}
\item Every correlation vector $p \in Q(S)$ for a scenario $S$ with finite inputs and outputs---as well as every subvector of such a vector---also belongs to $Q(G)$ for some finite exclusivity graph $G$. Conversely, every $p \in Q(G)$ for some finite exclusivity graph $G$ can be realized as a subvector of some correlation vector in $Q(S)$ for some scenario~$S$. Equivalently, if $\operatorname{Sub}\big(Q(S)\big)$ denotes the set of all subvectors of vectors in $Q(S)$, then
\begin{equation}
\bigcup_{S}\operatorname{Sub}\!\big(Q(S)\big) \;=\; \bigcup_{G} Q(G),
\end{equation}
where $S$ ranges over all measurement-contextuality scenarios\cite{Budroni:2022RMP} with finite inputs and outputs (not only Bell scenarios), and $G$ ranges over all finite exclusivity graphs. For the definition of measurement-contextuality scenario, see Appendix~\ref{app:contextuality}.

\item For every exclusivity graph $G$ that contains an imperfect induced subgraph, the quantum set $Q(G)$ has a curved boundary with rich structure\cite{CSW2014}. The collection of all such sets $Q(G)$ provides a detailed and, as we will show, distinctive signature of quantum theory.

\item Unlike Bell scenarios, exclusivity graphs do not specify how contexts are defined in terms of parties, measurement settings, and outcomes. They retain only the resulting exclusivity relations. Consequently, a single graph $G$ can arise from infinitely many distinct scenarios whose events have exclusivity graph $G$. If an exclusivity graph $H$ has $G$ as an induced subgraph, then every correlation on $G$ extends by zero on $V(H)\setminus V(G)$ to a correlation on $H$; under this identification, the correlation set of $G$ is contained in that of $H$. Embedding at the level of exclusivity graphs is simpler than embedding scenarios: it acts only on the graph data, whereas scenario embedding must also carry along and consistently preserve the entire context structure.

\item For an arbitrary Bell or measurement-contextuality scenario $S$, no universal algorithm can decide whether a correlation $p$ lies in the quantum set $Q(S)$, even when arbitrarily small errors in $p$ are allowed\cite{Ji:2021CACM,Vidick:2026,Lin:2026}. By contrast, for an arbitrary exclusivity graph $G$, whether a correlation $p$ is inside $Q(G)$ can be formulated as a semidefinite program\cite{CSW2014}.
\end{itemize}

We assume the standard definition of state, observable, compatibility (or joint measurability), and disturbance in general probabilistic theories\cite{Plavala:2023PR}; for convenience, these and other definitions are in Appendix~\ref{app:gpt}.

\begin{definition}[Ideal measurement] \label{def:one}
An ideal measurement of an observable $A$ is a measurement that
(i) is repeatable, in the sense that an immediate repetition on the same physical system yields the same outcome, and (ii) does not disturb any observable compatible with $A$.
\end{definition}

\begin{definition}[Correlation]
A correlation (or behavior or empirical model) is a table of probabilities for the outcomes of
joint measurements of compatible observables, produced in an experiment
in which every trial consists of such a joint measurement.
\end{definition}

As a solution to Problem~2, we propose the following principle.

\begin{principle}[Plenitude of ideal correlations (PIC)]\label{ax:pic}
Every correlation that is possible in a probabilistic theory in which every observable admits an ideal measurement is physically realizable.
\end{principle}

The paper is organized as follows.
Section~\ref{sec:PICG} proves that, for every exclusivity graph $G$,
the correlations satisfying PIC coincide exactly with the quantum correlation set $Q(G)$.
Building on this, Section~\ref{sec:PICS} shows that, for every measurement-contextuality scenario~$S$ (including every Bell scenario),
the correlations satisfying PIC coincide exactly with the quantum (commuting) correlation set $Q(S)$.
Section~\ref{sec:PICQT} then shows, as a further consequence, that PIC
singles out the Hilbert-space formalism and the Born rule of quantum theory.
Finally, Section~\ref{sec:discussion} discusses the scientific motivations for the interest in theories in which every observable can be measured ideally and some implications of our results.


\section{Proof that PIC singles out the quantum correlations $Q(G)$ for every exclusivity graph $G$} \label{sec:PICG}


\subsection{Mutually exclusive events}


\begin{definition}[Context]
A context is a set $\{x_i\}_{i=1}^N$ of jointly measurable observables.
\end{definition}

\begin{definition}[Event]
An event $(a_1,\ldots,a_N | x_1,\ldots,x_N)$ is an assignment of one outcome $a_i$ to each observable $x_i$, for $i=1,\ldots,N$, where $(x_1,\ldots,x_N)$ is a context.
\end{definition}

\begin{definition}[Event contained in an observable]
An observable $M$ contains an event $E$ if and only if $M$ has an outcome $m$ such that, for every state, $p(m)=p(E)$.
\end{definition}

\begin{definition}[Mutually exclusive events]
Two events $E$ and $F$ are mutually exclusive if and only if there exists an observable $M$ containing both $E$ and $F$ as distinct outcomes.
\end{definition}


\subsection{Compatibility graphs, exclusivity graphs and edge-colored exclusivity multigraphs} \label{sec:graphs}


A measurement-contextuality scenarios (and a Bell scenario) is characterized by a number of observables, the number of possible outcomes
of each observable, and the pairwise compatibility relations between the observables (see Appendix~\ref{app:contextuality}). These compatibility relations can be encoded by a graph.

\begin{definition}[Compatibility graph]
The compatibility graph of a set $V$ of observables is the graph
$G=(V,E)$ whose vertex set is $V$, and in which two distinct vertices
$u,v\in V$ are adjacent, written $u\sim v$, if and only if the
corresponding observables are compatible.
\end{definition}

In quantum theory, every finite compatibility graph can be realized. By contrast, in classical physics only {\em complete} compatibility graphs are realizable, because every observable is compatible with every other observable.

A correlation a measurement-contextuality or Bell scenario assigns probabilities to all possible events that can occur in that scenario. The relations of mutual exclusivity between any subset of events that can occur in that scenario can be encoded by a graph.

\begin{definition}[Exclusivity graph]
The exclusivity graph of a set $V$ of events is the graph $G=(V,E)$
whose vertex set is $V$, and in which two distinct vertices $u,v\in V$
are adjacent, written $u\sim v$, if and only if the corresponding events
are mutually exclusive.
\end{definition}

In both classical physics and quantum theory, every finite exclusivity graph can be realized by some events in some scenario.

The exclusivity graph does not encode which observables of the scenario are responsible for the relations of exclusivity. To also encode this information, we can use an edge-labelled multigraph.

\begin{definition}[Edge-colored exclusivity multigraph]
Let $S$ be a measurement-contextuality scenario with observables $\{O_j\}_{j=1}^N$, and let $V$ be a subset of the possible events occurring in
$S$. The edge-colored exclusivity multigraph of $V$ is the
edge-colored multigraph on a labelled vertex set
\begin{equation}
G=(V,E_1,\dots,E_N),
\end{equation}
whose vertex set is $V$, and two distinct events $u,v\in V$ are joined by an edge of color $j$ if and only if $u$ and $v$ are mutually exclusive outcomes of $O_j \in S$.
If $u$ and $v$ are events corresponding to mutually exclusive outcomes of two distinct
observables $O_j \neq O_k $ of $S$, then $u$ and $v$ are joined by two parallel edges, one of color $j$ and one of color $k$. Each event $u\in V$ is labeled by its observables and respective outcomes.
\end{definition}


\begin{figure}[h!]
\centering
\includegraphics[scale = 0.37]{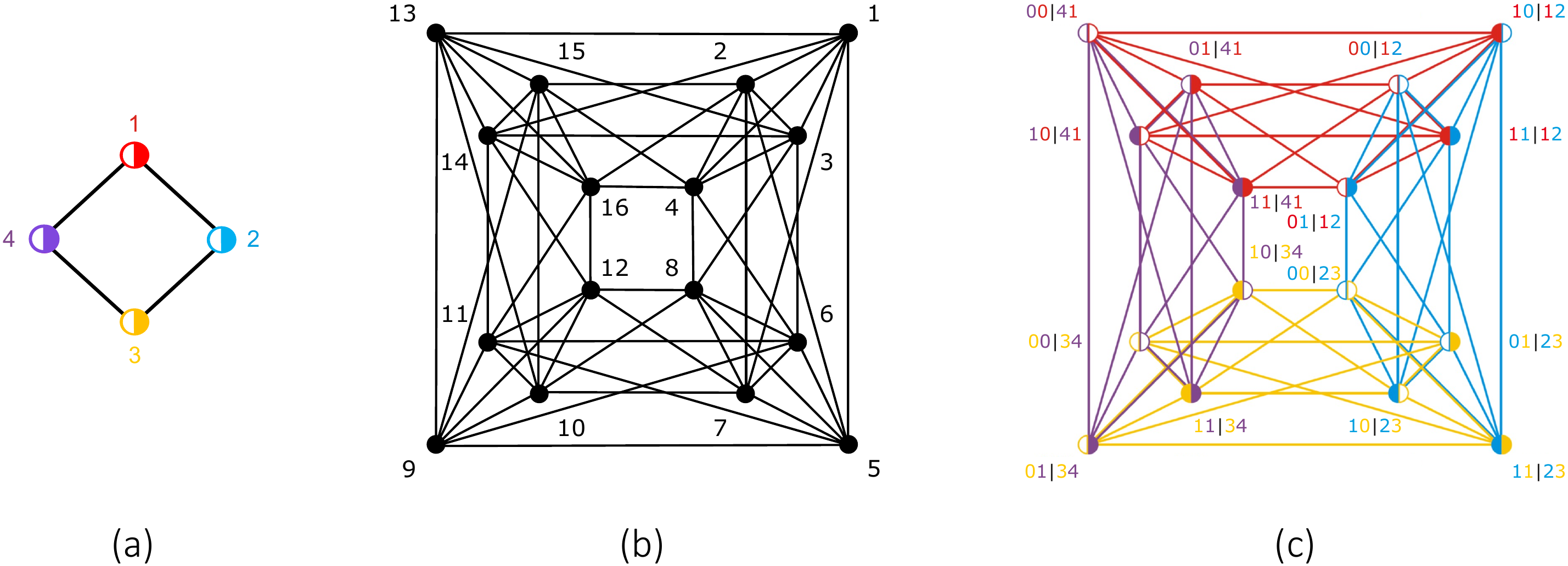}
\caption{(a)~The compatibility graph of the simplest scenario that allows non-classical (measurement-contextual) correlations. This scenario (graph) has four observables (vertices) and four contexts (cliques). It corresponds to the simplest Bell scenario where, in every round, Alice measures one of the observables $x \in \{1,3\}$, each with two outcomes $a \in \{0,1\}$, and Bob measures one of the observables $y \in \{2,4\}$, each with two outcomes $b \in \{0,1\}$\cite{Clauser:1969PRL}. The general measurement-contextuality scenario is called the four-cycle\cite{Araujo:2013PRA}. Each observable is represented by a vertex drawn as a circle, and a distinct color is assigned to each observable. Each circle is divided into two portions, each of which represents one of the two possible outcomes of that observable. Two vertices are connected by an edge if and only if the corresponding observables are compatible. 
(b)~The exclusivity graph of the $16$~possible events---i.e., measurements-and-outcomes combinations---that can occur in the simplest scenario. Each vertex represents an event, and two vertices (black circles) are connected by an edge if and only if the corresponding events are mutually exclusive.
(c)~The edge-colored exclusivity multigraph representing the $16$~possible events $(a,b|x,y)$ of the simplest scenario. Event~$(a,b|x,y)$ is labeled $ab|xy$, using the notations and colors introduced in~(a). Each event is represented by a vertex (colored circle) formed by combining the outcome portions of the observables that define it: the event $(a,b|x,y)$ is represented by a vertex consisting of the $a$-portion of observable $x$ and the $b$-portion of observable $y$. An edge of a given color connects two events if and only if those events have different outcomes for the observable represented by that color. If two events differ in their outcomes for more than one observable, then the corresponding vertices are joined by more than one colored edge.}
\label{fig:CEME}
\end{figure}


The simplest compatibility graph that allows non-classical correlations (see Appendix~\ref{app:contextuality}) is shown in Fig.~\ref{fig:CEME}(a), the exclusivity graph of its events is shown in Fig.~\ref{fig:CEME}(b), and the corresponding edge-colored exclusivity multigraph is shown in Fig.~\ref{fig:CEME}(c).

It is important to understand the differences between the exclusivity graph and the edge-colored exclusivity multigraph.

A correlation for the scenario $S$ whose compatibility graph is in Fig.~\ref{fig:CEME}(a) is a probability vector $p$ with $16$ probabilities, $p(a,b | x,y)$, one for each event $(a,b | x,y)$. Thus $p$ can be viewed as a vector assigned to the vertices of the edge-colored exclusivity multigraph in Fig.~\ref{fig:CEME}(c).

In $S$, each setting pair $(x,y)$ is a context, and for every context the four probabilities $p(a,b | x,y)$ over $a,b$ must sum to $1$. Hence the sum of all $16$ components of $p$ of a valid correlation for $S$ is $4$. 

The same $16$-component vector can also be assigned to the vertices of the unlabeled exclusivity graph $G$ in Fig.~\ref{fig:CEME}(b). However, $G$ retains only the mutual-exclusivity relations and discards colors, labels, and edge multiplicities. Consequently, the quantum set $Q(G)$ for the unlabeled exclusivity graph does not impose the context normalizations
$\sum_{a,b} p(a,b | x,y)=1$;
it imposes only non-negativity and the exclusivity constraints encoded in $G$ (equivalently, clique constraints $\sum_{v\in C}p_v\le 1$). Therefore, the all-zero vector $p'$ on the same $16$ vertices belongs to $Q(G)$, but it is not a valid full correlation vector for $S$.


\subsection{The exclusivity condition}


\begin{theorem}[\cite{Chiribella:2014,Cabello:2019PRA,Chiribella:2020PRR}]\label{sp}
If every observable can be measured ideally, then every finite set of pairwise compatible observables is jointly measurable.
\end{theorem}

\begin{corollary}
If every observable can be measured ideally, then, in the compatibility graph of the scenario, every maximal context is represented by a maximal clique (i.e., a subset $C$ of vertices in which every two distinct vertices are adjacent, and such that no outside vertex is adjacent to all the members of $C$).
\end{corollary}

\begin{theorem}[\cite{Chiribella:2014,Cabello:2019PRA,Chiribella:2020PRR}]\label{ep}
If every observable can be measured ideally, then for every state $\rho$ and every set $\{E_i\}$ of pairwise exclusive events produced by ideal measurements, 
\begin{equation}
\sum_i p_{\rho}(E_i) \le 1,
\end{equation}
where $p_{\rho}(E_i)$ denotes the probability of event $E_i$ in state $\rho$.
\end{theorem}

In other words,
\begin{equation}\label{epp}
\boxed{
\forall\,\text{state } \rho \text{ and } \forall\,\{E_i\}\ \text{pairwise exclusive events produced by ideal measurements:}\quad
\sum_i p_{\rho}(E_i)\le 1 .
}
\end{equation}

\begin{proof}
We prove Theorems~\ref{sp} and~\ref{ep} together. The proof is presented for a set $\{a,b,c\}$ of three pairwise exclusive events; the proof for any number of events is analogous.

Since $a$ and $b$ are mutually exclusive, there exists an observable $AB$ whose outcome set contains $a$ and $b$ as distinct outcomes (and possibly other outcomes). Similarly, since $a$ and $c$ are mutually exclusive, there exists an observable $AC$ whose outcome set contains $a$ and $c$; and since $b$ and $c$ are mutually exclusive, there exists an observable $BC$ whose outcome set contains $b$ and $c$.

Let $A$ be the coarse-graining of $AB$ with outcomes $\{a,\overline{a}\}$, where $\overline{a}$ denotes ``not $a$''; this is an observable compatible with $AB$. The same two-outcome observable $A$ is also the coarse-graining of $AC$ to $\{a,\overline{a}\}$, because outcome $a$ has a well-defined effect independent of the compatible observable in which it is embedded. Hence $A$ is compatible with $AC$ as well. Analogously, let $B$ be the coarse-graining of $AB$ (and of $BC$) with outcomes $\{b,\overline{b}\}$, and let $C$ be the coarse-graining of $AC$ (and of $BC$) with outcomes $\{c,\overline{c}\}$.

We now show that $A$, $B$, and $C$ are pairwise compatible. Since $A$ and $B$
are both coarse-grainings of $AB$, they are jointly measurable via $AB$;
similarly, $A$ and $C$ are jointly measurable via $AC$, and $B$ and $C$ via
$BC$. Hence $A$, $B$, and $C$ are pairwise compatible.

Since $A$, $B$, and $C$ can be measured ideally, we construct an ideal
measurement of an observable $D$ with outcomes $\{a,b,c,d\}$ as follows. First,
perform an ideal measurement of $A$; this disturbs neither $B$ nor $C$. If the
outcome of $A$ is $a$, we define the outcome of $D$ to be $a$. If the outcome of $A$ is $\overline{a}$, perform an ideal measurement of $B$ on the same physical system; this does not disturb $C$. If the outcome of $B$ is $b$, we define the outcome of $D$ to be $b$. If the outcome of $B$ is $\overline{b}$, perform an ideal measurement of $C$. If the outcome of $C$ is $c$, we define the outcome of $D$ to be $c$; if the outcome of $C$ is $\overline{c}$, we define the outcome of $D$ to be $d$. By construction, $D$ is an ideal measurement, and its outcomes $\{a,b,c,d\}$ are mutually exclusive. Therefore, for any state,
\begin{equation}
p_{\rho}(D=a)+p_{\rho}(D=b)+p_{\rho}(D=c)+p_{\rho}(D=d)=1.
\end{equation}
Hence
\begin{equation}
p_{\rho}(D=a)+p_{\rho}(D=b)+p_{\rho}(D=c)\le 1.
\end{equation}
By non-disturbance, $p_{\rho}(D=a)=p_{\rho}(a)$, $p_{\rho}(D=b)=p_{\rho}(b)$, and
$p_{\rho}(D=c)=p_{\rho}(c)$, so
\begin{equation}
p_{\rho}(a)+p_{\rho}(b)+p_{\rho}(c)\le 1.
\end{equation}
This proves Theorem~\ref{ep}.

Finally, $D$ is a common refinement of $A$, $B$, and $C$: coarse-graining $D$ to $\{a,\overline{a}\}$ yields $A$, to $\{b,\overline{b}\}$ yields $B$, and to
$\{c,\overline{c}\}$ yields $C$. Hence $A$, $B$, and $C$ are jointly measurable, which proves Theorem~\ref{sp}.
\end{proof}

Historically, the statement ``every set of pairwise compatible observables is jointly measurable'' has been called Specker's principle\cite{Cabello:2012XXX}.
Similarly, the condition $\sum_i p_{\rho}(a_i) \le 1$ for every state $\rho$ and every set $\{a_i\}$ of pairwise exclusive events has been called the exclusivity principle\cite{CSW2010,Cabello:2013PRL,Yan:2013PRL,AmaralPRA2014,amaral2018graph,Cabello:2019PRA,Chiribella:2020PRR}. Here we have shown that both follow as theorems from a single assumption: that every observable admits an ideal measurement.


\subsection{Independent experiments viewed as a joint experiment} \label{sec:iej}


\begin{definition}[Independent experiments]\label{def:ind}
Let $A$ and $B$ be experiments with state spaces $\Omega_A,\Omega_B$ and
observable sets $\mathcal O_A,\mathcal O_B$, respectively, that can be performed in the same trial. We say that $A$ and $B$ are independent if, for all states $\rho_A\in\Omega_A$, $\rho_B\in\Omega_B$ and all observables $O_A=\{A_a\}_{a\in \mathcal A} \in \mathcal O_A$, $O_B=\{B_b\}_{b\in \mathcal B} \in \mathcal O_B$, chosen without reference to each other's outcomes,
\begin{equation}\label{eq:ind}
p(a,b|\rho_A,\rho_B,O_A,O_B)=p(a |\rho_A,O_A)\,p(b|\rho_B,O_B)\qquad\forall(a,b)\in\mathcal A\times\mathcal B.
\end{equation}
\end{definition}

\begin{definition}[Joint experiment]\label{def:joint}
Let $A,B$ be independent experiments that can be performed in the same trial. The joint experiment $(A,B)$ is specified by:
\begin{enumerate}
\item \emph{Preparations.} $\Omega_{(A,B)}:=\mathrm{conv}\{\rho_A\otimes\rho_B\}$, where $\rho_A\otimes\rho_B$ is the preparation of $\rho_A$ and $\rho_B$ in the same trial. Here $\otimes$ is the product of functions on pairs of states, not a tensor product of vector spaces: it is well defined on equivalence classes by \eqref{eq:ind}, and its factors are recovered with the unit effects.
\item \emph{Protocols.} For an experiment $A$, $\PP(A)\subseteq\OO_A$ denotes the observables of $A$ that may be used as stages of a protocol; every observable of $A$ must be a post-processing of one of them. For an elementary experiment, $\PP(A):=\OO_A$ (all its observables). If $M=\{E_i\}_{i\in I}\in\PP(A)$ and $N_i=\{F_{ij}\}_{j\in J_i}\in\PP(B)$ for every $i$, ``perform $M$ on $A$; if $i$ occurs, perform $N_i$ on $B$'' is a protocol $(M,(N_i))$ with outcomes $(i,j)$ and effects $E_i\otimes F_{ij}$, where $E\otimes F$ is the effect $\rho_A\otimes\rho_B\mapsto\rho_A(E)\rho_B(F)$, extended affinely. The same holds with $A,B$ exchanged, giving $(N,(M_j))$. The case $N_i\equiv N$ is the \emph{parallel} measurement of $M$ and $N$, with the effects of $M\boxtimes N:=\{E_i\otimes F_j\}$. Let $\PP(A,B)$ denote the set of protocols for the joint experiment $(A,B)$.
\item \emph{Observables.} The observables of $(A,B)$ are the classical post-processings of elements of $\PP(A,B)$. The unit effect is $u_A\otimes u_B$.
\item \emph{Statistics.} On $\rho_A\otimes\rho_B$, $p(i,j)=p(i | \rho_A,M)\,p(j | \rho_B,N_i)$: within the branch $i$, the run is the parallel run $M\boxtimes N_i$, so by \eqref{eq:ind} its statistics depend only on the operations along that branch.
\item \emph{Instruments.} If $(i,j)$ occurs, the post-measurement state is $\rho'_{A,i}\otimes\rho'_{B,ij}$, where $\rho'_{A,i}$ and $\rho'_{B,ij}$ are the local post-measurement states; the extension to mixtures follows by linearity.
\end{enumerate}
\end{definition}

\begin{lemma}[Ideal protocols]
Let $A,B$ be independent experiments that are performed in the same trial, and let every observable of $A$ (of $B$) be a post-processing of an element of $\mathcal{P}(A)$ (of $\mathcal{P}(B)$), all admitting ideal measurements.
If~$M\in\mathcal{P}(A)$ and $N_i\in\mathcal{P}(B)$ are ideal measurements $\sfm,\sfn_i$, then the procedure ``perform $\sfm$; if $i$ occurs, perform $\sfn_i$'' is an ideal measurement of the protocol $S=(M,(N_i))$ in $(A,B)$; likewise with $A,B$ exchanged. In particular, the parallel measurement of ideal measurements is an ideal measurement of $M\boxtimes N$.
\end{lemma}

\begin{proof}
Let
\begin{equation}
S=(M,(N_i)),\qquad
M=\{M_i\}_{i\in I},\qquad
N_i=\{N_{ij}\}_{j\in J_i},
\end{equation}
and let $\sfS$ denote the procedure ``perform $\sfm$; if $i$ occurs, perform $\sfn_i$''. We show that $\sfS$ is an ideal measurement of $S$.

\medskip
{\em 1. Repeatability.}
Suppose that outcome $(i,j)$ occurs. By Definition~\ref{def:joint}(5), the post-measurement state is
\begin{equation}
\rho'_{A,i}\otimes \rho'_{B,ij}.
\end{equation}
By the repeatability property of ideal measurements, on $\rho'_{A,i}$ the measurement $\sfm$ gives $i$ with certainty, and on $\rho'_{B,ij}$ the measurement $\sfn_i$ gives $j$ with certainty. By independence, the probability of repeating $(i,j)$ is therefore $1$. Hence $\sfS$ is repeatable for $S$.

{\em 2. Non-disturbance.}
Let $P$ be an observable jointly measurable with $S$. Then $P$ and $S$ are post-processings of a common protocol $T\in\mathcal{P}(A,B)$. Since post-processing preserves undisturbed statistics, it is enough to prove that $\sfS$ does not change the statistics of $T$. By affinity, we may take a product initial state $\rho_A\otimes\rho_B$.

Write
\begin{equation}
T=(X,(Y_x)),\qquad
X=\{X_x\}_{x\in X}\in\mathcal{P}(A),\quad
Y_x\in\mathcal{P}(B),
\end{equation}
and
\begin{equation}
S=(M,(N_i)),\qquad
M=\{M_i\}_{i\in I}\in\mathcal{P}(A),\quad
N_i\in\mathcal{P}(B).
\end{equation}
There are two cases, according to whether the first stage of $T$ is on $A$ or on $B$. 

We will use the following consequence of repeatability: if an outcome $z$ of an ideal measurement $Z$ can occur, then there is a state in which $z$ is certain. Indeed, take the post-measurement state $\rho'_z$; by repeatability, $p(z|\rho'_z,Z)=1$.

\medskip
{\em Case 1: $T=(X,(Y_x))$ starts on $A$.}
Since $S$ is a post-processing of $T$, after summing over $j$ and allowing randomization, the first-stage outcome $i$ of $S$ is a post-processing of the first-stage outcome $x$ of $T$. Thus $M$ is a post-processing of $X$, written
\begin{equation}
M\preceq X,
\end{equation}
with a possibly randomized post-processing kernel $\pi(i|x)$:
\begin{equation} \label{uno}
p(i|\rho_A,M)
=
\sum_x \pi(i|x)\,p(x|\rho_A,X).
\end{equation}
On the post-measurement state $\rho'_{A,i}$, repeatability of $\sfm$ gives
\begin{equation}
p(i|\rho'_{A,i},M)=1.
\end{equation}
Using \eqref{uno},
\begin{equation}
1
=
\sum_x \pi(i|x)\,p(x|\rho'_{A,i},X).
\end{equation}
Since $\pi(i|x)\le 1$ and the weights $p(x|\rho'_{A,i},X)$ sum to $1$, 
\begin{equation}
\pi(i|x)=1
\end{equation}
for every $x$ with $p(x|\rho'_{A,i},X)>0$. Thus, in every branch $x$ that can occur after outcome $i$, the outcome $i$ is certain.

Fix such a branch $x$. Since $x$ can occur on $\rho'_{A,i}$, by the consequence of repeatability recalled above there exists a state $\rho_A^x$ in which $x$ is certain for $X$. In that state, by \eqref{uno},
\begin{equation}
p(i|\rho_A^x,M)=\pi(i|x)=1.
\end{equation}
Hence in that state both $x$ and $i$ are certain. The post-processing relation between $S$ and $T$ then reduces to
\begin{equation}
N_i\preceq Y_x.
\end{equation}
By non-disturbance, the ideal measurement $\sfm$ does not disturb the statistics of $X$, and the ideal measurement $\sfn_i$ does not disturb the statistics of $Y_x$ in the corresponding branch. Explicitly, the statistics of $T$ after the non-selective $\sfS$ are
\begin{equation}
\sum_i p_i\,p(x|\rho'_{A,i},X)\,p(y|\tau_i,Y_x),
\end{equation}
where $p_i=p(i|\rho_A,M)$ and $\tau_i$ is the non-selective post-state of $\sfn_i$. On each branch that contributes, $p(y|\tau_i,Y_x)=p(y|\rho_B,Y_x)$, and $\sum_i p_i\,p(x|\rho'_{A,i},X)=p(x|\rho_A,X)$; this gives the statistics of $T$ without $\sfS$. Finally, by independence and Definition~\ref{def:joint}(4), the branches factorize. Hence the statistics of $T$ after measurement $\sfS$ are the same as those of $T$ without~$\sfS$.

\medskip
{\em Case 2: $T=(Y,(X_y))$ starts on $B$.}
Since $S$ is a post-processing of $T$, for every $y$ that can occur, by the consequence of repeatability recalled above there exists a state $\rho_B^y$ in which $y$ is certain for $Y$. Evaluating the post-processing relation in the state $\rho_A\otimes\rho_B^y$, the first-stage outcome $i$ of $S$ is a randomized post-processing of $X_y$, so
\begin{equation}
M\preceq X_y.
\end{equation}
Next evaluate the post-processing relation in the post-measurement state $\rho'_{A,i}$, where $i$ is certain by repeatability of $\sfm$.
Then the second-stage outcome $j$ of $S$ is a randomized post-processing of $Y$, so
\begin{equation}
N_i\preceq Y.
\end{equation}
By non-disturbance, $\sfm$ does not disturb $X_y$, and $\sfn_i$ does not disturb $Y$. Again by independence and Definition~\ref{def:joint}(4), the branches factorize, and the statistics of $T$ after the non-selective measurement $\sfS$ are the same as those of $T$ without~$\sfS$.

\medskip
In both cases, $\sfS$ does not disturb $T$, and therefore does not disturb any post-processing $P$. So $\sfS$ is non-disturbing.

Consequently, $\sfS$ is an ideal measurement of $S$. The case with $A,B$ exchanged is proved identically after swapping the labels. Taking $N_i\equiv N$ gives the parallel measurement $M\boxtimes N$.
\end{proof}

\begin{definition}[OR product, also called co-normal or disjunctive product]
Let $G_1=(V_1,E_1)$ and $G_2=(V_2,E_2)$ be two graphs. The OR product
$G_1 \lor G_2$ is the graph with vertex set
\begin{equation}
V(G_1 \lor G_2) = V_1 \times V_2,
\end{equation}
where two distinct vertices $(u,v)$ and $(u',v')$ are adjacent if and
only if
\begin{equation}
u u' \in E_1 \quad\text{or}\quad v v' \in E_2.
\end{equation}
\end{definition}

Let the events in experiment $A$ have exclusivity graph $G=(V,E)$: vertices correspond to events $A_u$, and $u\sim w$ if and only if $A_u,A_w$ are distinct outcomes of one observable of $A$. On $A_1,\dots,A_N$ (copies of $A$) put $G_{\mathbf u}:=A_{u_1}\otimes\dots\otimes A_{u_N}$, $\mathbf u\in V^N$. Let $\Ex_{\PP}$ be the graph on $\{G_{\mathbf u}\}$ in which two events are adjacent if and only if they are distinct outcomes of an element of $\PP_N$, and $\Ex$ the one for all observables of $(A_1,\dots,A_N)$; $\Ex_{\PP}\subseteq\Ex$.

\begin{definition}[Mutually independent copies]
Let $A_1,\dots,A_N$ be copies of an experiment $A$. They are mutually independent if they can be performed in the same trial and, for every $k=2,\dots,N$, the joint experiment $(A_1,\dots,A_{k-1})$ and $A_k$ are independent. Here
$(A_1,\dots,A_k):=((A_1,\dots,A_{k-1}),A_k)$ is defined recursively by Definition~\ref{def:joint}, with states $\mathrm{conv}\{\rho_1\otimes\cdots\otimes\rho_k\}$, protocols $\mathcal P_k:=\mathcal P(A_1,\dots,A_k)$, and observables their post-processings. In particular, for every product preparation $\rho_1\otimes\cdots\otimes\rho_N$ and observables $O_1,\dots,O_N$ chosen without reference to each other's outcomes, $p(a_1,\dots,a_N)=\prod_kp(a_k | \rho_k,O_k)$, and by \eqref{eq:ind} the observable of one copy may be chosen as a function of the outcomes already obtained on the others.
\end{definition}

\begin{proposition}\label{prop:or}
Let $A_1,\dots,A_N$ be mutually independent copies of $A$.
\begin{enumerate}
\item[(a)] $\Ex_{\PP}=G^{\vee N}$.
\item[(b)] $\Ex\supseteq G^{\vee N}$. Equality holds if, for every pair of
non-mutually-exclusive events $A,A'$, there is a state $\rho$ such that
$\rho(A)=1$ and $\rho(A')>0$.
\end{enumerate}
\end{proposition}
\begin{proof}
($\supseteq$ in (a)) Induction on $N$; $N=1$ is the definition of $G$. Let $k$ be such that $u_k\sim u'_k$. If $k\le N-1$, by induction there is $W\in\PP_{N-1}$ having $G_{u_{<N}},G_{u'_{<N}}$ as distinct outcomes. Take $Y_x$ an observable of $A_N$ containing $A_{u_N}$ if $x$ is the outcome $G_{u_{<N}}$, containing $A_{u'_N}$ if $x$ is the outcome $G_{u'_{<N}}$ (every event is an outcome of some observable), and arbitrary otherwise. The protocol $(W,(Y_x))$ has $G_{\mathbf u},G_{\mathbf u'}$ as distinct outcomes. If $k=N$, measure first an observable of $A_N$ with $A_{u_N},A_{u'_N}$ as distinct outcomes, then a parallel observable of the first $N-1$ copies containing $G_{u_{<N}}$ (resp.\ $G_{u'_{<N}}$): a protocol $(Y,(X_y))\in\PP_N$.

($\subseteq$ in (a)) Let $G_{\mathbf u}\ne G_{\mathbf u'}$ be distinct outcomes of $(X,(Y_x))$ (the other order is symmetric). By uniqueness of factors, $X_x=G_{u_{<N}}$, $X_{x'}=G_{u'_{<N}}$ and $Y_{xy}=A_{u_N}$, $Y_{x'y'}=A_{u'_N}$. If $x\ne x'$ they are distinct outcomes of $X\in\PP_{N-1}$, so $u_{<N}\sim u'_{<N}$ in $G^{\vee(N-1)}$ by induction; if $x=x'$ then $y\ne y'$ are distinct outcomes of $Y_x$ and $u_N\sim u'_N$. In both cases $\mathbf u\sim\mathbf u'$ in $G^{\vee N}$.

($\supseteq$ in (b)) From (a), since $\Ex_{\PP}\subseteq\Ex$.

($\subseteq$ in (b)) Recall that if $A, A'$ are not mutually exclusive, then there exists a state $s$ such that $s(A)=1$ and $s(A')>0$. Suppose $u_k\not\sim u'_k$ for all $k$. Then, pick $\rho_k$ with $\rho_k(A_{u_k})=1$, $\rho_k(A_{u'_k})>0$. For $w=\rho_1\otimes\dots\otimes\rho_N$, $w(G_{\mathbf u})=1$ and $w(G_{\mathbf u'})>0$, so they cannot be distinct outcomes of one observable.
\end{proof}


\subsection{Multiple copies as one joint experiment reveal invalid correlations}


A correlation $p$ may appear valid yet fail to be valid because two copies of it, i.e., $p\otimes p$, fail to satisfy the exclusivity condition \eqref{epp}.
For example, consider the correlation
\begin{equation}
p=\Bigl(\tfrac12,\tfrac12,\tfrac12,\tfrac12,\tfrac12\Bigr)
\end{equation}
on the five vertices of $G=C_5$ (i.e., a pentagon). This $p$ is not a valid correlation. To see this, form the product correlation $p\otimes p$ on $G\vee G$, whose component at the vertex $(i,j)$ is
\begin{equation}
(p\otimes p)_{(i,j)}=p_i p_j=\tfrac14 .
\end{equation}
If the vertices of the pentagon 
$G$ are numbered sequentially, then
the five vertices 
\begin{equation} \label{example}
(1,1),\ (2,4),\ (3,2),\ (4,5),\ (5,3)
\end{equation}
of $G\vee G$ form a clique; hence the corresponding events are pairwise exclusive, and $p\otimes p$ assigns probability $\tfrac14$ to each of them. The sum of these probabilities is therefore
\begin{equation}
5\cdot \tfrac14=\tfrac54>1,
\end{equation}
which violates the exclusivity condition \eqref{epp}. Hence $p\otimes p$ is not a valid correlation; consequently, $p$ itself is not a valid correlation\cite{Cabello:2013PRL}.

It can also occur that detecting that a correlation is invalid requires considering not two independent copies but an arbitrarily large number of them. 
The difficulty is that detecting a violation of \eqref{epp} requires searching over cliques in the $k$-fold OR graph product $G^{\vee k}$, which has $n^{k}$ vertices---i.e., a graph whose size grows exponentially with the number of copies $k$---and the associated clique-finding problems are NP-hard.

So the question becomes:
\begin{quote}
\textbf{Problem 3.} Given a graph $G$, characterize the set of correlations $p$ on $G$ such that, for every $k\ge 1$, the $k$-fold product correlation $p^{\otimes k}$ on the $k$-fold OR graph product $G^{\vee k}$ satisfies the exclusivity condition \eqref{epp}.
\end{quote}

In ~\cite{CSW2014}, we observed that, for $k=1$, this set is precisely the fractional stable set polytope of $G$.

\begin{definition}[Fractional stable set polytope\footnote{The concept of the fractional stable set polytope emerged from the effort to approximate the stable set polytope by a tractable linear relaxation, and it became a central object in polyhedral combinatorics and perfect graph theory. Its development spans the early 1970s to the late 1980s.}]
Let $G=(V,E)$. The fractional stable set polytope of $G$, also known as the fractional vertex packing polytope or the clique-constrained stable set polytope, is
\begin{equation}
\QSTAB(G) = \left\{ p\in\mathbb{R}_{\ge 0}^{V} :\,\sum_{i \in C} p_i \le 1 \ \text{for every clique } C \subseteq V \right\}.
\end{equation}
\end{definition}

Therefore, the set of correlations solution of Problem~3 can be written, formally, as
\begin{equation}
\QSTAB^\infty(G)
:=
\bigl\{p:\ p^{\otimes k}\in \QSTAB(G^{\vee k})
\text{ for every } k\ge 1\bigr\}.
\end{equation}

The difficulty is that, in general, deciding membership in $\QSTAB^\infty(G)$ is computationally hard.


\subsection{Two complementary experiments reveal invalid correlations}


\begin{definition}[Weighted Lovász theta number]
Let $G=(V,E)$ be a finite simple graph and let
$w\in\mathbb{R}_{\ge 0}^{V}$ be a non-negative weight vector.
The weighted Lovász theta number $\vartheta(G,w)$ is defined by the
semidefinite program
\begin{equation}
\vartheta(G,w)
=
\max\left\{
\sum_{v\in V} w_v\,p_v
:\;
p\in\mathbb{R}^{V},\ 
U\in\mathbb{S}^{V},\ 
\begin{pmatrix}1 & p^{\top}\\ p & U\end{pmatrix}\succeq 0,\ 
U_{vv}=p_v\ \ \forall v\in V,\ 
U_{uv}=0\ \ \forall uv\in E(G)
\right\},
\end{equation}
where $\mathbb{S}^{V}$ denotes the set of real symmetric matrices indexed by 
$V$.

For $w=\mathbf 1$, one recovers the ordinary Lovász theta number
$\vartheta(G)$\footnote{The Lovász theta number $\vartheta(G)$ was introduced by László Lovász in 1979\cite{Lovasz79}.
The original aim of the Lovász theta number was to estimate and upper-bound the Shannon capacity of a graph, i.e., the zero-error capacity of a noisy communication channel whose confusable symbol pairs are encoded by the edges of the graph. The theta number subsequently became important because it has the following properties: (i) It can be approximated to arbitrary precision in polynomial time via semidefinite programming. (ii) It satisfies the sandwich inequality
\begin{equation}
\alpha(G)\le \vartheta(G)\le \chi(\overline G),
\end{equation}
so $\vartheta(G)$ is sandwiched between the clique number
$\omega(\overline G)=\alpha(G)$ and the chromatic number
$\chi(\overline G)$ of the complement graph $\overline G$.
Both the clique number and the chromatic number are NP-hard to compute exactly in general (and are also hard to approximate).}.
\end{definition}

\begin{definition}[Theta body\footnote{The theta body was introduced by Gr\"otschel, Lov\'asz, and Schrijver in 1986\cite{Grotschel:1986JCTB} as the tractable convex relaxation of the independent-set polytope derived from the Lovász theta number.}]
Let $G=(V,E)$. The theta body of $G$ is\cite{Grotschel:1986JCTB}[p.~337]
\begin{equation}
\operatorname{TH}(G)
=
\left\{
p\in\mathbb{R}_{\ge 0}^{V}
:\;
w^{\top}p\le \vartheta(G,w)
\quad
\text{for every } w\in\mathbb{R}_{\ge 0}^{V}
\right\},
\end{equation}
where $\vartheta(G,w)$ is the weighted theta number, $w$ is a weight vector, and $w^{\top}$ its transpose.
\end{definition}

\begin{theorem} \label{valid}
Every $p \in \operatorname{TH}(G)$ satisfies that, for every $k\ge 1$,
$p^{\otimes k} \in \QSTAB^{\vee k}(G)$.
\end{theorem}

\begin{proof}
If
$p\in \operatorname{TH}(G)$ then, for every $k\ge 1$, $p^{\otimes k}\in \operatorname{TH}(G^{\vee k})$.
Since
\begin{equation}
\operatorname{TH}(G^{\vee k})\subseteq \QSTAB(G^{\vee k}),
\end{equation}
we conclude that, for any $k \ge 1$,
\begin{equation}
p^{\otimes k}\in \QSTAB^{\vee k}(G).
\end{equation}
\end{proof}

\begin{definition}[Complement of a graph]
Let $G=(V,E)$ be a simple graph. The complement of $G$ is the graph
$\overline{G}=(V,\overline{E})$,
where
\begin{equation}
\overline{E}
=
\bigl\{\{u,v\} : u,v\in V,\ u\neq v,\ \{u,v\}\notin E\bigr\}.
\end{equation}
Equivalently, two distinct vertices are adjacent in $\overline{G}$ if and only if they are not adjacent in $G$.
\end{definition}

\begin{theorem}[\cite{AmaralPRA2014}]
\label{th:atc1}
Assume that every $p \in \operatorname{TH}(G)$ is a physically realizable
correlation for the exclusivity graph $G$. Then, the exclusivity condition
\eqref{epp} rules out every correlation $q$ for $\overline{G}$ such that
$q \notin \operatorname{TH}(\overline{G})$.
\end{theorem}

\begin{proof}
Let $\{p_i\}$ be a correlation for the $N$-vertex exclusivity graph $G$,
and let $\{q_i\}$ be a correlation for the $N$-vertex exclusivity graph
$\overline{G}$. Assume that $G$ and $\overline{G}$ correspond to events of
independent experiments. Consider a clique of size~$N$ in the exclusivity
graph $G \wedge \overline{G}$ representing joint events; such a clique is constructed explicitly in \eqref{example}. Let $\{r_i\}$ be the correlation for that clique. Since the events are independent, the probabilities factor as
\begin{equation}
r_i = p_i q_i .
\end{equation}
Applying the exclusivity condition \eqref{epp} to this clique:
\begin{equation}\label{eq:Eprinciple}
\sum_{i} p_i q_i \le 1 .
\end{equation}

By assumption, every $p \in \operatorname{TH}(G)$ is a physically realizable correlation for $G$. The theta body can also be written as follows\cite{Grotschel:1986JCTB}[p.~337]:
\begin{equation}\label{eq:THdef}
\operatorname{TH}(G)
= \bigl\{ p \in \mathbb{R}_{\ge 0}^{V} :
\vartheta(\overline{G}, p) \le 1 \bigr\}.
\end{equation}
By the duality between $\operatorname{TH}(G)$ and
$\operatorname{TH}(\overline{G})$ encoded in \eqref{eq:THdef}, for a given
$p$, inequality \eqref{eq:Eprinciple} holds for all
$q \in \operatorname{TH}(\overline{G})$ if and only if
$p \in \operatorname{TH}(G)$. Consequently, if
$q \notin \operatorname{TH}(\overline{G})$, then \eqref{eq:Eprinciple}
fails for some $p \in \operatorname{TH}(G)$, i.e.\, such a $q$ is ruled out
by~\eqref{epp}. Equivalently, if $p \notin \operatorname{TH}(G)$, there
exists $q \in \operatorname{TH}(\overline{G})$ with
$p \otimes q \notin \QSTAB(G \wedge \overline{G})$.
\end{proof}

\begin{theorem}[\cite{AmaralPRA2014}]
\label{th:atc2}
If $G$ is a self-complementary exclusivity graph, then the exclusivity condition \eqref{epp} alone excludes every correlation that is not contained in $\operatorname{TH}(G)$.
\end{theorem}

\begin{proof}
Let $p$ be a correlation for $G$ with
$p \notin \operatorname{TH}(G)$. By Theorem~\ref{th:atc1} (more
precisely, by the dual statement established at the end of its proof),
there exists $q \in \operatorname{TH}(\overline{G})$ such that
\begin{equation}\label{eq:exclusion}
\sum_{i \in V(G)} p_i q_i > 1.
\end{equation}
Since $G$ is self-complementary, let
$\pi : \overline{G} \to G$ be an isomorphism, regarded as a
permutation of the index set $V(G)$. Define $q'$ by
$q'_{\pi(i)} = q_i$; then $q' \in \operatorname{TH}(G)$. Because
$\pi$ merely permutes the indices and the sum is invariant under
relabeling,
\begin{equation}
\sum_{i \in V(G)} p_i q'_i
= \sum_{i \in V(G)} p_{\pi(i)}\, q_i
= \sum_{i \in V(G)} p_i q_i
> 1.
\end{equation}
Thus $p$ and $q'$ are two correlations on the same graph $G$
whose product violates inequality~\eqref{epp}; equivalently, $p$
cannot be a correlation allowed by the exclusivity
condition~\eqref{epp} on $G$. Since $p \notin \operatorname{TH}(G)$
was arbitrary, the exclusivity condition \eqref{epp} alone excludes every correlation not contained in $\operatorname{TH}(G)$.
\end{proof}

Hence, for every self-complementary exclusivity graph $G$, the set of correlations achievable in any physical theory in which every observable admits an ideal measurement is contained in $\operatorname{TH}(G)$. Moreover, for
such $G$, assuming the principle of plenitude of ideal correlations (PIC), every
correlation in $\operatorname{TH}(G)$ is physically realizable.


\subsection{Result for arbitrary graphs}


\begin{lemma}\label{lem:in}
Let $H$ be a finite graph and let $G=H[V(G)]$ be the induced subgraph of
$H$ on a vertex subset $V(G)\subseteq V(H)$. Then, the projection of
$\operatorname{TH}(H)$ onto the coordinates indexed by $V(G)$ is exactly
$\operatorname{TH}(G)$:
\begin{equation}
\operatorname{TH}(H)\big|_G = \operatorname{TH}(G).
\end{equation}
\end{lemma}

\begin{proof}
Assume $H$ is finite and $G=H[V(G)]$ is the induced subgraph of $H$ on a subset $V(G)\subseteq V(H)$. Write
\begin{equation}
\operatorname{TH}(H)\big|_G
:=
\Bigl\{p\in\mathbb{R}_{\ge 0}^{V(G)}:\ \exists\,q\in\mathbb{R}_{\ge 0}^{V(H)\setminus V(G)}
\text{ with }(p,q)\in\operatorname{TH}(H)\Bigr\}.
\end{equation}
We use the characterization \eqref{eq:THdef}.
For $p\in\mathbb{R}_{\ge 0}^{V(G)}$, let $(p,0)$ denote its extension by zero to $V(H)$.

Since $G$ is induced in $H$, we have $\overline G=\overline H[V(G)]$. Two elementary properties of the weighted Lovász theta function hold:
\begin{enumerate}
\item \emph{Zero extension:}
\begin{equation}
\vartheta(\overline H,(p,0))=\vartheta(\overline G,p).
\end{equation}
Indeed, any orthonormal representation of $\overline G$ extends to one of $\overline H$ by assigning the zero-weight vertices fresh mutually orthogonal vectors orthogonal to all previous ones; conversely, restricting an orthonormal representation of $\overline H$ to $V(G)$ gives one of $\overline G$.
\item \emph{Monotonicity:} if $0\le r\le s$ coordinatewise on $V(H)$, then
\begin{equation}
\vartheta(\overline H,r)\le \vartheta(\overline H,s).
\end{equation}
\end{enumerate}

Now prove both inclusions.

\emph{First inclusion: $\operatorname{TH}(H)|_G\subseteq \operatorname{TH}(G)$.}
Let $p\in\operatorname{TH}(H)|_G$. Then, there exists $q\ge 0$ on $V(H)\setminus V(G)$ such that
\begin{equation}
\vartheta(\overline H,(p,q))\le 1.
\end{equation}
Since $(p,0)\le (p,q)$, monotonicity and zero extension give
\begin{equation}
\vartheta(\overline G,p)
=
\vartheta(\overline H,(p,0))
\le
\vartheta(\overline H,(p,q))
\le 1.
\end{equation}
Hence $p\in\operatorname{TH}(G)$.

\emph{Second inclusion: $\operatorname{TH}(G)\subseteq \operatorname{TH}(H)|_G$.}
Let $p\in\operatorname{TH}(G)$. Then, $\vartheta(\overline G,p)\le 1$. By zero extension,
\begin{equation}
\vartheta(\overline H,(p,0))
=
\vartheta(\overline G,p)
\le 1,
\end{equation}
so $(p,0)\in\operatorname{TH}(H)$. Therefore, $p\in\operatorname{TH}(H)|_G$.

Combining both inclusions,
\begin{equation}
\operatorname{TH}(H)\big|_G=\operatorname{TH}(G).
\end{equation}
\end{proof}

\begin{theorem}[\cite{Cabello:2019PRA}]
If $G$ is an arbitrary exclusivity graph, PIC implies that every correlation in $\operatorname{TH}(G)$ is physically realizable.
\end{theorem}

\begin{proof}
The following construction shows that every finite graph $G=(V,E)$ is an
induced subgraph of a self-complementary graph. Let $A,B,C,D$ be four copies
of $V$. On $A$ and $D$, put a copy of $G$; on $B$ and $C$, put a copy
of $\overline{G}$. Add all edges between $A$ and $B$, all edges between
$B$ and $C$, and all edges between $C$ and $D$, and no other
inter-block edges. The resulting graph $H$ has $4|V|$ vertices and contains
$G$ as the induced subgraph on $A$.

The graph $H$ is self-complementary. Indeed, the block permutation
\begin{equation}
A\mapsto B,\qquad B\mapsto D,\qquad C\mapsto A,\qquad D\mapsto C
\end{equation}
is an isomorphism from $H$ to $\overline{H}$.

Since $H$ is self-complementary, PIC implies that every correlation in $\operatorname{TH}(G)$ is physically realizable. By Lemma~\ref{lem:in},
\begin{equation}
\operatorname{TH}(H)|_A = \operatorname{TH}(G),
\end{equation}
which proves the theorem.
\end{proof}

\begin{theorem}[\cite{CSW2014}]
For every exclusivity graph $G$, the quantum set of correlations generated by ideal measurements is $\operatorname{TH}(G)$.
\end{theorem}

\begin{proof}
In Dirac's notation, the theta body $\operatorname{TH}(G)$ is
\begin{equation}
\operatorname{TH}(G)
=
\left\{
p\in\mathbb{R}^{V}:
\begin{array}{l}
\exists\,\{|u_i\rangle\}_{i\in V},\ \exists\,|\psi\rangle
\text{ such that } \langle u_i|u_i\rangle=1,\ 
\langle \psi|\psi\rangle=1,\\[2pt]
\langle u_i|u_j\rangle=0 \text{ for all } ij\in E(G),
\text{ and }
p_i=|\langle u_i|\psi\rangle|^2\ \forall i\in V
\end{array}
\right\}.
\end{equation}
\end{proof}

Hence, for every exclusivity graph $G$,
\begin{itemize}
\item The set of correlations achievable in any physical theory in which every observable admits an ideal measurement is contained in $\operatorname{TH}(G)$.
\item Assuming the principle of plenitude of ideal correlations (PIC), every
correlation in $\operatorname{TH}(G)$ is physically realizable.
\item PIC singles out exactly the set of quantum correlations generated by ideal measurements.
\end{itemize} 


\section{Proof that PIC singles out the quantum set $Q(S)$ for every scenario $S$} \label{sec:PICS}


Here we show that Problem~2 implies Problem~1: a physical principle that
singles out the quantum correlations for ideal measurements on any
exclusivity graph also singles out the quantum correlations in every Bell
scenario.

\begin{theorem} \label{th:tri}
If every observable admits an ideal measurement, then, for any Bell scenario $S$, every correlation achievable in $S$ is achievable using ideal measurements alone. Equivalently, the correlation set obtained with arbitrary measurements coincides with the correlation set obtained with ideal measurements.
\end{theorem}

\begin{proof}
By hypothesis, each local observable admits an ideal measurement. Therefore, replacing every local measurement in the Bell scenario by such an ideal measurement leaves the correlation unchanged. Hence, every correlation achievable with arbitrary measurements is achievable with ideal measurements alone.
\end{proof}

Recall the discussion in Sec.~\ref{sec:graphs} on the relationship between the
edge-colored exclusivity multigraph $M_S$ and its corresponding exclusivity
graph $G_S$.

Our result that, for every exclusivity graph $G$, the set of correlations
achievable in any physical theory in which every observable admits an ideal
measurement is contained in $\operatorname{TH}(G)$, together with
Theorem~\ref{th:tri}, implies that, for any Bell scenario $S$, the set of
correlations assigned to the vertices of the edge-colored exclusivity
multigraph $M_S$ is contained in $\operatorname{TH}(G_S)$.

\begin{theorem} \label{th:sa}
If for every measurement-contextuality scenario $S$, the allowed correlations for every exclusivity graph $G$ are elements of $\operatorname{TH}(G)$, then the set of effects characterizing an observable (that can be ideally measured) must be a set of orthogonal projectors that sum to the identity operator.
\end{theorem}

\begin{proof}
Suppose that $S$ consists of only one observable with a set $R$ of $N$ possible outcomes. Then $G_S$ is a complete graph on $N$ vertices, and $M_S$ is a complete multi-graph on $N$ vertices whose edges are all single edges of the same color, since there is only one observable. In this case, a correlation $p \in \operatorname{TH}(G_S)$ must satisfy an extra condition to be a valid correlation for the vertices of $M_S$: the sum of all probabilities in $p$ must be $1$. Since the vectors $\{|u_i\rangle\}_{i \in R}$ are orthogonal, this extra condition implies that, for every unit vector $|\psi\rangle$,
\begin{equation}
\sum_{i \in R} |\langle u_i|\psi\rangle|^2
= \sum_{i \in R} \langle \psi |u_i\rangle \langle u_i|\psi\rangle
= \left\langle \psi \left| \left( \sum_{i \in R} |u_i\rangle \langle u_i| \right) \right| \psi \right\rangle = 1.
\end{equation}
Since this holds for every unit vector $|\psi\rangle$, it follows that
\begin{equation}
\sum_{i \in R} |u_i\rangle \langle u_i| = \1,
\end{equation}
so the projectors $\{|u_i\rangle \langle u_i|\}_{i \in R}$ are orthogonal and sum to the identity operator $\1$.
\end{proof}

A family of orthogonal projectors that sum to the identity is exactly a projection-valued measure, and a self-adjoint operator is the standard way to encode such a family once the outcomes are labeled by real numbers. The equivalence is the finite-dimensional spectral theorem.

\begin{theorem}\label{zp2}
If for every measurement-contextuality scenario $S$, the valid correlations for every exclusivity graph~$G$ are elements of $\operatorname{TH}(G)$, then compatible observables (that can be ideally measured) are represented by commuting self-adjoint operators.
\end{theorem}

\begin{proof}
Let $X$ and $Y$ be compatible observables with outcome sets $A$ and $B$, respectively. A joint measurement of $X$ and $Y$ is equivalent to a measurement of their common refinement $XY$. Consider the scenario $S$ corresponding to $XY$. Its exclusivity graph $G_S$ is the complete graph on $M\times N$ vertices, where $M=|A|$ and $N=|B|$. For each event $(a,b)\in A\times B$, let
\begin{equation}
R_{ab}=|u_{(a,b)}\rangle\langle u_{(a,b)}|
\end{equation}
be the corresponding projector.

By hypothesis, the joint correlation belongs to $\operatorname{TH}(G_S)$. Theorem~\ref{th:sa} implies that the projectors $R_{ab}$ form an orthogonal resolution of the identity:
\begin{equation}
R_{ab}R_{a'b'}=\delta_{aa'}\delta_{bb'}R_{ab},
\qquad
\sum_{a,b}R_{ab}=\1.
\end{equation}

Define the marginals
\begin{equation}
P_a:=\sum_{b\in B}R_{ab},
\qquad
Q_b:=\sum_{a\in A}R_{ab}.
\end{equation}
Using the orthogonality of the $R_{ab}$, we obtain
\begin{equation}
P_a^\dagger=P_a,\qquad P_a^2=P_a,\qquad P_aP_{a'}=0\ (a\neq a'),
\qquad \sum_a P_a=\1,
\end{equation}
and similarly
\begin{equation}
Q_b^\dagger=Q_b,\qquad Q_b^2=Q_b,\qquad Q_bQ_{b'}=0\ (b\neq b'),
\qquad \sum_b Q_b=\1.
\end{equation}
Thus $\{P_a\}$ and $\{Q_b\}$ are projection-valued measures for $X$ and $Y$, respectively.

Moreover,
\begin{equation}
P_aQ_b
=\Bigl(\sum_{b'}R_{ab'}\Bigr)\Bigl(\sum_{a'}R_{a'b}\Bigr)
=R_{ab},
\end{equation}
and similarly $Q_bP_a=R_{ab}$. Hence
\begin{equation}
R_{ab}=P_aQ_b=Q_bP_a,
\end{equation}
so the joint projectors factor through the marginals and, in particular,
\begin{equation}
[P_a,Q_b]=0
\qquad \forall a,b.
\end{equation}
Therefore,
\begin{equation}
X=\sum_{a\in A}a\,P_a,
\qquad
Y=\sum_{b\in B}b\,Q_b
\end{equation}
are represented by commuting self-adjoint operators, with spectral integrals in the continuous case. Since $X$ and $Y$ were arbitrary compatible ideally measured observables, the claim follows.
\end{proof}

Combining that for every measurement-contextuality scenario $S$, the only physically realizable correlations for every exclusivity graph $G$ are elements of $\operatorname{TH}(G)$---as we have shown in Sec.~\ref{sec:PICG}---, and Theorems \ref{th:sa} and \ref{zp2}, we conclude the following: 

Plenitude of ideal correlations (PIC) implies that, for any measurement-contextuality scenario $S$ (including any Bell scenario), every correlation $p=p(a,\ldots,c|x,\ldots,z)$ that satisfies the following conditions is physically realizable: 
\begin{equation}
p= \langle \psi | \Pi_{a|x} \cdots \Pi_{c|z} | \psi \rangle,
\end{equation}
where $|\psi\rangle$ is a unit vector in a complex\footnote{The use of complex numbers is a matter of convenience\cite{Dagmar}: any alternative mathematical representation that yields the same sets of correlations is acceptable.} Hilbert space $\mathcal H$, 
\begin{equation}
\Pi_{a|x}\Pi_{a'|x}=\delta_{aa'}\Pi_{a|x},
\quad
\sum_a \Pi_{a|x}=\1_A,\quad
\ldots, \quad
\Pi_{c|z}\Pi_{c'|z}=\delta_{cc'}\Pi_{c|z},
\quad
\sum_c \Pi_{c|z}=\1_C,
\end{equation}
and
\begin{equation}
[\Pi_{a|x},\Pi_{c|z}]=0
\qquad \forall a,c,
\end{equation}
and similarly for any two distinct compatible observables. That is, projectors belonging to distinct observables commute. Here, $\1_A$ (respectively, $\1_C$) is the identity operator on the subspace $\mathcal H_A$ (respectively, $\mathcal H_C$) of $\mathcal H$.

Hence PIC leads to the commuting quantum set rather than to the tensor-product quantum set.  For quantum systems described by finite-dimensional Hilbert spaces both sets are equal. For infinite-dimensional Hilbert spaces, the commuting quantum set contains the tensor-product quantum set, and the inclusion is strict in some scenarios\cite{Ji:2021CACM,Vidick:2026}. For a discussion of the physical interpretation of this strict inclusion, see Ref.~\cite{cabello2023logicalpossibilitiesphysicsmipre}.


\section{PIC singles out quantum theory} \label{sec:PICQT}


By now, it should be clear that the PIC also singles out the Hilbert-space
formalism and the Born rule of quantum theory. More precisely, it tells us why
pure states are represented by rays in a Hilbert space (with representative
vector $\lvert\psi\rangle$), why the atomic extremal effects are represented
by rank-one projectors $\lvert u\rangle\langle u\rvert$ (with $\lVert u\rVert=1$), and why the corresponding probabilities are given by the Born rule
\begin{equation}
\langle\psi\vert u\rangle\langle u\vert\psi\rangle
=
\lvert\langle u\vert\psi\rangle\rvert^2 .
\end{equation}

This Hilbert-space--Born-rule structure admits, and gives a natural account of, further features of quantum theory: observables may be represented by POVMs,
and, reciprocally---via Naimark's dilation theorem---every POVM can be realized as a projective measurement (a PVM) on a larger Hilbert space. Similarly, every
mixed state can be viewed as the reduced state (marginal) of a pure state on a
larger Hilbert space.

By Wigner's theorem, every bijection of the state space that preserves all transition probabilities $|\langle \phi | \psi \rangle|^2$ is induced by a unitary or antiunitary operator on Hilbert space, unique up to a phase. Hence, if each time-translation operator~$T_t$ preserves transition probabilities---for instance, because the system is closed---then each $T_t$ is induced by a unitary or antiunitary operator.

The group law $T_{t+\tau}=T_tT_\tau$ rules out antiunitary time translations. Indeed, for any $t$,
\begin{equation}
T_t = T_{t/2}T_{t/2},
\end{equation}
and the square of a unitary or antiunitary operator is unitary; therefore, every $T_t$ is unitary.

Finally, assuming strong continuity of $t\mapsto T_t$, Stone's theorem gives a unique self-adjoint generator $H$ such that
\begin{equation}
T_t = e^{-iHt/\hbar},
\end{equation}
and the Schr\"odinger equation
\begin{equation}
i\hbar\,\partial_t\psi = H\psi
\end{equation}
holds on the domain of $H$.

In addition, although the PIC is dimension-agnostic, PIC also implies that systems described by infinite-dimensional Hilbert spaces are necessary. The reason is that some allowed correlations are attainable only in infinite dimensions: if only finite-dimensional Hilbert spaces were allowed, then the full plenitude of allowed correlations would not be achievable, even in Bell scenarios with a finite---and small---number of inputs and outputs\cite{Coladangelo:2020NC,Pauwels2026,Coladangelo2026}.


\section{Discussion} \label{sec:discussion}


\subsection{Why theories of observales admitting ideal measurements?}


Why physical theories in which every observable admits an ideal measurement are interesting? 

\begin{itemize}
\item \emph{The notion of an ideal measurement is operational}: it is specified by operational criteria---repeatability and minimal disturbance.

\item \emph{Ideal measurements are minimally disturbing}. By the definition of compatibility (see Appendix~\ref{app:gpt}), $A$ and $B$ are compatible if and only if they admit a joint measurement; equivalently, if $A$ and $B$ are incompatible, then \emph{every} measurement of $A$ disturbs $B$ for some state. Hence, if \emph{some} measurement of $A$ leaves $B$ undisturbed for all states, then $A$ and $B$ must be compatible. The converse, however, does not hold: compatibility of $A$ and $B$ does not guarantee that a measurement of $A$ exists that leaves $B$ undisturbed. The ideal measurement of $A$ is precisely the one that leaves every observable compatible with $A$ undisturbed.

\item \emph{Ideal measurements are sufficient for maximal predictability without disturbance.} Consider an experiment consisting of sequential measurements of mutually compatible observables. If each measurement is ideal, then earlier measurements do not disturb later ones, for any initial state. Consequently the joint outcome statistics of the entire sequence are order-independent and can be predicted by the theory from the initial state alone. In this sense, ideal measurements are what make such a theory maximally predictive.

\item \emph{Ideal measurements play a central role in general Bayesian theories}; this is discussed extensively in Ref.~\cite{Chiribella:2020PRR}.

\item \emph{Ideal measurements are fundamental for physical \emph{probabilistic} theories of the Einstein--Podolsky--Rosen (EPR) type\cite{Einstein:1935PR}}, which we
define as those probabilistic theories\footnote{Einstein, Podolsky, and Rosen themselves were skeptical of probabilistic theories.} in which, for every observable and every outcome, there exists some state in which that outcome can be predicted with certainty without disturbing compatible observables.

This definition is obtained from the EPR criterion in two steps. Einstein,
Podolsky, and Rosen identify the following as an essential feature of every physical theory:
\begin{quote}
If, without in any way disturbing a system, one can predict with certainty
(i.e., with probability equal to unity) the value of a physical quantity, then there exists an element of physical reality corresponding to this physical quantity; and every element of physical reality must have a counterpart in the physical theory.
\end{quote}
First, we drop the commitment to an ``element of reality.'' Second, we turn the predictability condition from a criterion into a requirement on the theory: for every observable and every outcome, the antecedent must be satisfiable in some state. The condition that every outcome be certain in some state and the existence of an ideal measurement for every observable are related as follows. Existence of a certainty state for each outcome is necessary for idealness. If every observable admits an ideal measurement, then any outcome that is certain in a given state can be predicted with certainty by measuring that observable, without disturbing compatible observables. This is the role ideal measurements play in EPR-type probabilistic theories. 

\item \emph{In both classical physics and quantum theory, every observable admits an ideal measurement.} What differs between the two is not the availability of ideal measurements but the extent to which observables are compatible: in classical physics all observables are jointly measurable, whereas in quantum theory they are not.
\end{itemize}


\subsection{An explanation of the world}


The principle of plenitude---sometimes called the totalitarian principle and formulated as ``everything allowed by the laws of nature must actually exist''\cite{Kragh2019XXX}---has appeared in physics in different variants\cite{Kragh2019XXX}. Here, we use it with a specific meaning: ``everything allowed'' refers to every correlation that is mathematically possible in theories in which every observable admits an ideal measurement. We use it for a specific purpose: to understand quantum theory\cite{Feynman:1967}.

In what sense can we now claim to better understand quantum theory?

First, we can now \emph{derive} quantum theory from a \emph{physical} principle. We are finally in a situation in which ``one makes sense of all of the formalism on the basis of first principles''\cite{Grinbaum}.

Second, we can now identify \emph{under which epistemological assumption} the principle of plenitude of ideal correlations selects quantum theory: every observable admits an ideal measurement. Adopting theories of this type is a scientific choice, and like other choices in science it is justified by what it buys us---here, greater predictive power.

Third, unlike a mere reconstruction---whose ``explanatory power ($\ldots$) is a power of explanation of where the structure of the theory comes from; not necessarily a power of explanation by the theory, of the real world''\cite{Grinbaum}---our derivation also offers an explanation of the world.

Plenitude demands that no physical restriction prevent the saturation of the mathematical possibilities: every possible correlation is to be realized. This gives precise content to Max Born's conjecture that the success of quantum theory may be due to an \emph{ontological}
``nonexistence of inner properties that condition a particular outcome''\cite{Born:1926ZPb}. More precisely, what plenitude requires is the absence of inner properties that would rule out some otherwise allowed correlation.

The ontological lack alone, however, does not fix quantum theory: plenitude of ideal correlations is supplemented by the epistemological assumption that every observable is measured ideally. Quantum theory should therefore be understood as the consequence of the nonexistence of restrictive inner properties together with the adoption of probabilistic theories in which every observable admits an ideal measurement.

The parallel with the aether---the medium thought to fill space and carry light waves---is instructive. The absence of the aether has experimentally observable signatures when examined from inertial frames: the speed of light is the same in all of them. Likewise, the nonexistence of inner properties that condition a particular outcome has experimentally observable signatures when examined using theories of ideal observables: non-locality and contextuality saturate then the allowed possibilities.
We regard this as the most compelling reading of our results.


\subsection{An interpretation of quantum theory}


Existing interpretations of quantum theory\cite{cabellomap} fall essentially into three broad types:

\begin{itemize}
\item Those that treat quantum theory as a set of operational rules---a powerful recipe for predicting outcome probabilities---without explaining where the Hilbert-space--Born-rule formalism comes from.

\item Those that wrap the formalism in a global interpretive story invoking
wholeness, non-separability, or a perpetually branching universe---a story neither required by nor derived from the formalism, and one that explains neither the formalism nor, in particular, the Born rule.

\item Those that modify the fundamental dynamics of quantum theory,
claiming that the standard dynamics is merely an approximation, in order to
explain the emergence of definite outcomes and the Born rule.
\end{itemize}

None of these approaches is free of obscurities and tensions.

Our derivation, by contrast, explains why the formalism has the exact mathematical structure it does, where the Born rule comes from, and why the non-local and contextual correlations have the limits they do.

Once quantum theory is derived from a physical principle, the demand for an interpretation loses much of its force: it is unnecessary to add an interpretation on top of the theory---the physical principle itself does all the explanatory work.


\subsection{What remains to be understood}


We have not yet finished understanding quantum theory. Several questions remain open: what is operationally required for an experiment to realize all the possibilities allowed by the Hilbert-space--Born-rule formalism---as well as what happens when the conditions for such an experiment are not met. We will address these questions in future work.


\section*{Acknowledgements}


This paper presents refined and improved versions of earlier results. In this sense, it can be
regarded as a review. It also adds new intermediate
theorems, simplifies some proofs, and sharpens others. We hope this clarify the main message and make it more accessible.

The need of such improvements became clear after years of conversations. For those I have to acknoledge specially B\'arbara Amaral, Costantino Budroni, Giulio Chiribella, Miquel Feixas, Pietro Gambazza, Otfried G\"uhne, Matthias Kleinmann, Jan-\AA{}ke Larsson, Markus P.\ M\"uller, Miguel Navascu\'es, Tommaso Pedroni, Martin Plávala, Simone Severini, Marcelo Terra Cunha, Frank Wilczek, Andreas Winter, and Zhen-Peng Xu. We additionally thank Tommaso Pedroni for his help with Sec.~\ref{sec:iej}.

This work was supported by the AEI/MICIU Project ``Quantum Correlations for Quantum Information and Computation'' (QCQIC) (Project No.~PID2025-174467NB-I00).


\newpage

\appendix


\section{Contextuality} \label{app:contextuality}


We review the concept of measurement contextuality (also called Kochen--Specker contextuality~\cite{Budroni:2022RMP}) and explain why it is not only a natural generalization of Bell non-locality but also an essential perspective for understanding quantum theory.

A frequent academic exercise is to wonder\cite{Mermin:1985PT}:
\begin{quote}
How do you think Einstein would have reacted to Bell's theorem?
\end{quote}

I suspect how Einstein would have reacted to the formulation of Bell's theorem by Suppes and Zanotti\cite{Suppes:1981SYN}, and Fine\cite{Fine:1982PRL}:

\begin{definition}[Bell non-locality]
A correlation $p$ in a Bell experiment violates a Bell inequality (i.e., is Bell non-local) if and only if there is no global probability distribution over the outcomes of all observables whose marginals reproduce the probability distributions in~$p$.
\end{definition}

I suspect Einstein would have been surprised by the very possibility that no global probability distribution over the outcomes of all observables need exist. And Einstein's surprise would be reasonable, since such a possibility is rare.

\begin{theorem}[Vorob'ev\cite{Vorobev1962}]
Given compatible marginals, a global joint probability distribution consistent with all of them is guaranteed to exist whenever the compatibility graph is chordal (i.e., contains no induced cycle of length greater than three).
\end{theorem}

The basic non-chordal graphs---those containing no proper induced non-chordal subgraph---are precisely the cycles of length greater than three (the square, the pentagon, etc.). The square corresponds to the simplest Bell scenario.
However, apart from the square, none of these cycles arises as the compatibility graph of a Bell scenario.

\begin{quote}
{\bf Problem 3.} How can {\em all} correlations not ruled out by Vorob'ev's theorem be produced operationally, and in what type of experiment?
\end{quote}


\begin{figure}[h!]
\centering
\includegraphics[scale = 0.19]{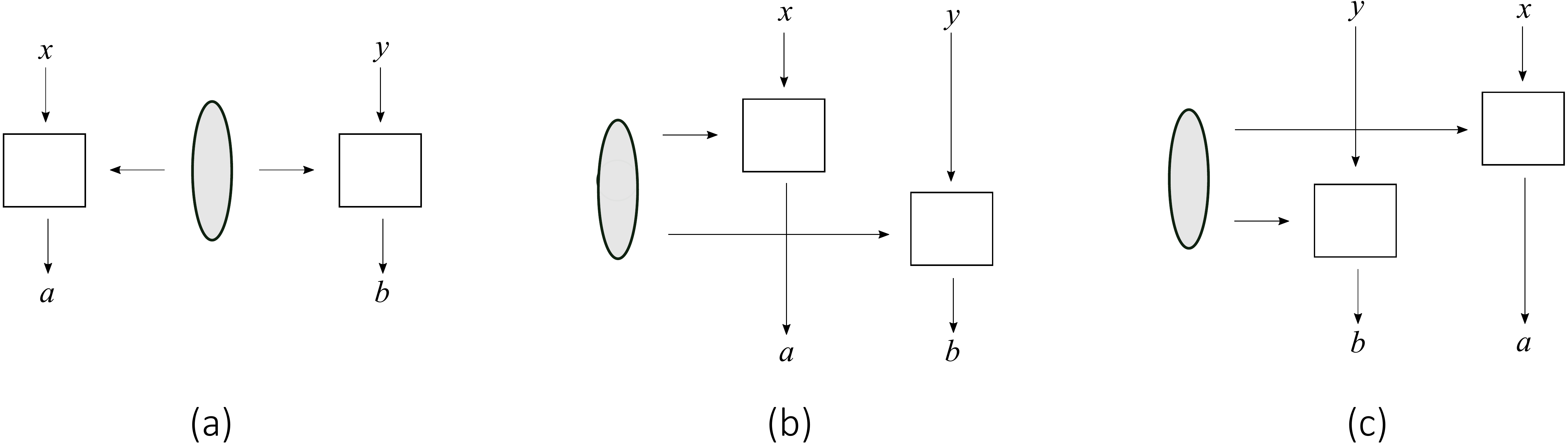}
\caption{Three equivalent temporal orderings of the same Bell experiment.}
\label{slb}
\end{figure}


When we represent a Bell experiment, we typically use a scheme like the one in
Fig.~\ref{slb}(a). However, the fact that one party's choice of input is
spacelike separated from the other party's outcome implies that we can also
represent the same experiment as in Fig.~\ref{slb}(b), where Alice measures
first and her measurement does not disturb the statistics of any observable that
Bob may measure later, or as in Fig.~\ref{slb}(c), where Bob measures first and
his measurement does not disturb the statistics of any observable that Alice may
measure later.

The equivalence of these three temporal orderings means that, in a Bell
experiment, the joint outcome statistics do not depend on which measurement is
considered to occur first. This motivates considering sequential measurements on
a single physical system in which the order of the measurements is required to
be irrelevant. Such experiments are the natural single-system analogue of Bell
experiments with equivalent temporal orderings.

Therefore, the experiments that extend Bell experiments and allow us to observe correlations not ruled out by Vorob'ev's theorem are sequential experiments on a single physical system. In such experiments, the order of the measurements must be irrelevant: for any state and any pair of outcomes,
\begin{equation}
p(a,b|x,y) = p(b,a|y,x).
\end{equation}
Thus the observables $x$ and $y$ must be jointly measurable, and measuring one
must not disturb the outcome statistics of any observable compatible with the
one being measured.

Moreover, for the outcome of a measurement to allow one to predict with
certainty that repeating the same measurement will yield the same
outcome---regardless of which compatible observable is measured between the two
measurements---the measurement must be repeatable.

These two features---non-disturbance of any compatible observable and
repeatability---are the operational defining characteristics of ideal
measurements (see Definition~\ref{def:one}).

Therefore, the experiments we are looking for are those in which, for a given state $\rho$, each trial consists of a sequence of ideal measurements of compatible observables. We call such an experiment a measurement-contextuality experiment.

\begin{definition}[Measurement contextuality]
A correlation $p$ in a measurement-contextuality experiment violates a non-contextuality inequality (i.e., is measurement-contextual) if and only if there is no global probability distribution over the outcomes of all observables whose marginals reproduce the probability distributions in~$p$.
\end{definition}

\begin{definition}[Measurement-contextuality scenario]
A measurement-contextuality scenario is specified by a set of observables, their respective sets of outcomes, and the joint-measurability relations among these observables.
\end{definition}

\begin{definition}[Bell scenario]
A Bell scenario is specified by the number of spatially separated parties, for each party the set of incompatible observables they may choose to measure, and for each such observable the set of possible outcomes.
\end{definition}

Why is measurement contextuality so important? Because the possibilities left open by Vorob'ev's theorem are not merely mathematical: they are realized in nature.

\begin{theorem}[{\cite{Xu:2019PRA}}]
For every non-chordal compatibility graph (i.e., one containing an induced cycle of length greater than three), quantum theory allows for measurement-contextual correlations.
\end{theorem}

These contextual correlations, moreover, are a resource for quantum computational advantage\cite{Budroni:2022RMP}[Sec.~VI A]\cite{Bharti:2026}.


\section{General probabilistic theories} \label{app:gpt}


\begin{definition}[Preparation]
A preparation is a reproducible experimental procedure involving an interaction
between a physical system and a device. Formally, it is represented as a process
with no input system and a physical system as output.
\end{definition}

\begin{definition}[Measurement]
A measurement is an interaction between a physical system and a measuring
device that produces a stable record consisting of exactly one outcome from a
well-defined set of possible outcomes, which may be labeled by real numbers.
The measurement may leave a physical system available for further measurements.
Formally, a measurement is a process that takes a physical system as input and
produces as output an outcome and, possibly, a physical system.
\end{definition}


\begin{figure}[t!]
\centering
\includegraphics[scale = 0.35]{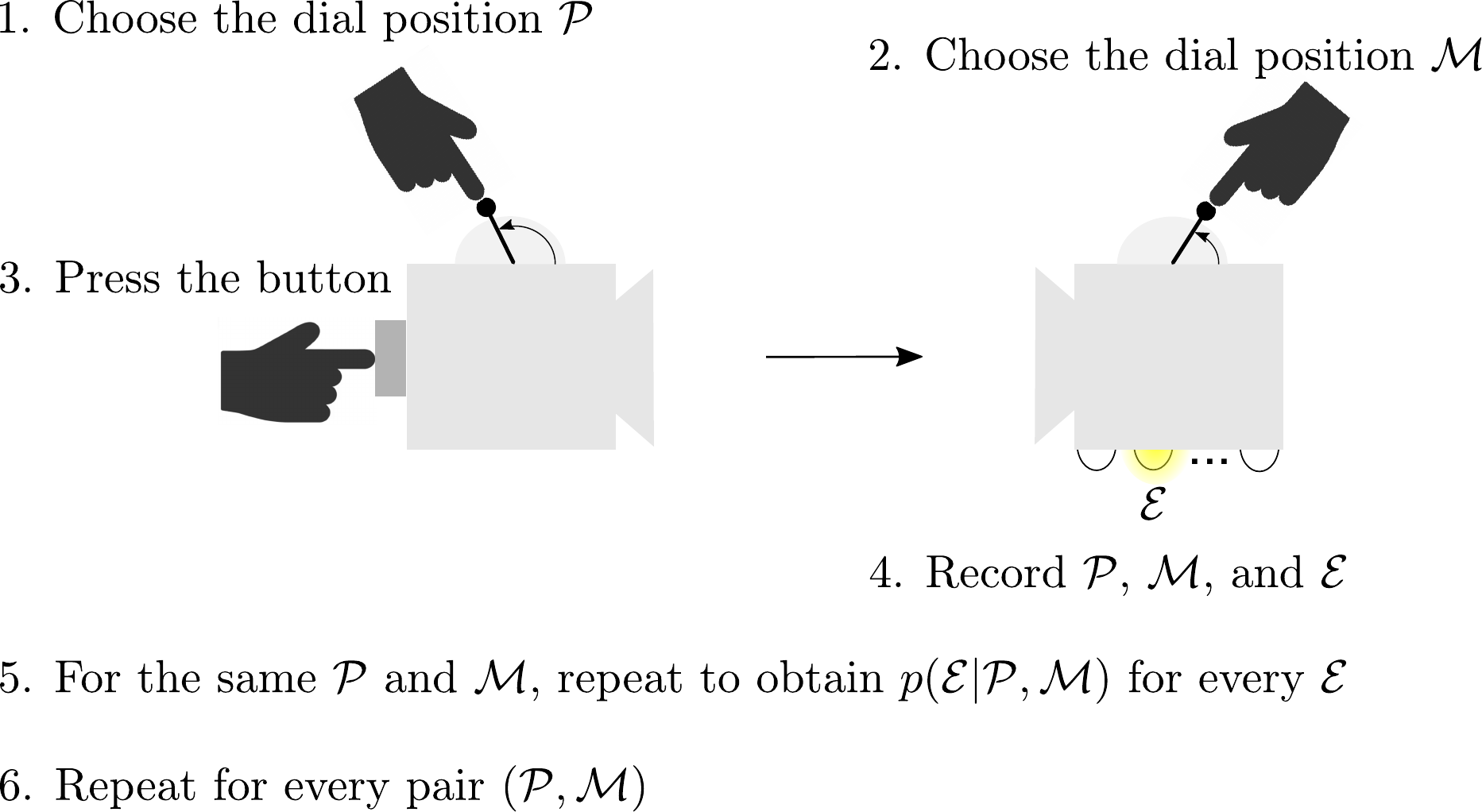}
\caption{\emph{The operational scenario of general probabilistic theories.} Each
preparation corresponds to a dial position $\mathcal{P}$ on the left-hand box;
each measurement corresponds to a dial position $\mathcal{M}$ on the right-hand
box. After selecting $\mathcal{P}$ and $\mathcal{M}$, the experimenter presses
the button on the preparation box, releasing a physical system. This causes
exactly one of the bulbs on the measurement box to flash---that is, the outcome
$\mathcal{E}$ of measurement $\mathcal{M}$ for preparation $\mathcal{P}$. For
fixed $\mathcal{P}$ and $\mathcal{M}$, by repeatedly pressing the button we
obtain a probability distribution $p(\mathcal{E} | \mathcal{P},\mathcal{M})$. Repeating this process for all pairs $(\mathcal{P},\mathcal{M})$, we obtain a collection of such distributions. This information allows us to partition the preparations and measurements into equivalence classes. Two preparations are considered equivalent if they yield the same probability distribution over outcomes for every measurement; each such equivalence class is called a \emph{state}. Similarly, two measurements are equivalent if they yield the same probability distribution over outcomes for every preparation; each such equivalence class is called an \emph{observable}.}
\label{states_obs}
\end{figure}


Let us now consider a set of preparations and a set of measurements that can be
performed on the physical systems produced by those preparations. For simplicity,
we suppose that each distinct preparation can be labeled by a single parameter
$\mathcal{P}\in[0,1]$ and each distinct measurement by a single parameter
$\mathcal{M}\in[0,1]$, as in Fig.~\ref{states_obs}.

We now combine each preparation $\mathcal{P}$ with each measurement
$\mathcal{M}$ and repeat the experiment many times to obtain the probability of
each possible outcome $\mathcal{E}$ of each measurement $\mathcal{M}$ for each
preparation $\mathcal{P}$. These probabilities must be reproducible: each time
we prepare $\mathcal{P}$ and perform the measurement $\mathcal{M}$, we should
obtain the same probabilities for the outcomes.

\begin{definition}[State]
If two preparations $\mathcal{P}$ and $\mathcal{P}'$ (e.g., corresponding to two
different positions of the dial of the left-hand box in
Fig.~\ref{states_obs}) lead to the same probability distribution of outcomes for
every measurement, then we say that the two preparations are equivalent. Each
equivalence class is called a state.
\end{definition}

\begin{definition}[Observable]
If two measurements $\mathcal{M}$ and $\mathcal{M}'$ (e.g., corresponding to two
different positions of the dial of the right-hand box in
Fig.~\ref{states_obs}) lead to the same probability distribution of outcomes for
every preparation, then we say that the two measurements are equivalent. Each
equivalence class is called an observable.
\end{definition}

\begin{definition}[Effect]
An effect is the mathematical representation of a measurement outcome: it maps
each state to the probability of that outcome.
\end{definition}

\begin{definition}[Instrument]
An instrument is the mathematical representation of a measurement, including the
corresponding post-measurement state: it maps each input state to a family of
subnormalized post-measurement states indexed by the possible outcomes, such that
the normalization of each member gives the probability of the corresponding
outcome. The sum of the members is the post-measurement state when the outcome is
ignored.
\end{definition}

Neither states nor effects can be specified without reference to the other: a
state is defined by the probabilities it assigns to all effects, and an effect is
defined by the probabilities it assigns to all states. Specifically, a state
$\rho$ is a function on the set of all effects $\mathcal{F}$,
\begin{equation}
\rho: \mathcal{F} \rightarrow [0,1];
\end{equation}
the state $\rho$ gives the probability $p(a|\rho)$ for any effect
$a\in\mathcal{F}$. Each state is a catalog of probabilities for all possible
effects.

Similarly, an effect $a$ is a function on the set of all states $\Omega$,
\begin{equation}
a: \Omega \rightarrow [0,1];
\end{equation}
the effect $a$ gives the probability $p(a|\rho)$ for any state
$\rho\in\Omega$. Each effect is a catalog of probabilities for all possible
states. For any state $\rho$ and any effect $a$, there is a unique probability
$p(a|\rho)$ that $a$ occurs given that the state is $\rho$.

An observable $A$ with outcome set $\mathcal{A}$ is therefore a set of effects
$\{A_a\}_{a\in\mathcal{A}}$ such that $\sum_a A_a=u$, where $u$ is the unit
effect, i.e., the unique effect assigning probability $u(\rho)=1$ for every state
$\rho$. 

\begin{definition}[Jointly measurable (or compatible) observables]
Two observables $A=\{A_a\}_{a\in\mathcal{A}}$ and $B=\{B_b\}_{b\in\mathcal{B}}$
are jointly measurable or compatible if there exists a joint
observable $J=\{J_{(a,b)}\}_{(a,b)\in\mathcal{A}\times\mathcal{B}}$ such that,
for all states $\rho$ and all $a\in\mathcal{A}$, $b\in\mathcal{B}$,
\begin{subequations}
\begin{align}
A_a(\rho) &= \sum_{b} J_{(a,b)}(\rho),\\
B_b(\rho) &= \sum_{a} J_{(a,b)}(\rho),
\end{align}
\end{subequations}
where each $J_{(a,b)}$ is an effect, i.e.,
\begin{equation}
0 \le J_{(a,b)}(\rho) \le 1
\quad\text{for all }\rho,
\qquad
\sum_{(a,b)} J_{(a,b)}(\rho) = u(\rho)=1.
\end{equation}

Equivalently, $A$ and $B$ are jointly measurable if and only if there exists a
common refinement $C=\{C_c\}_{c\in\mathcal{C}}$ and functions
$f\colon\mathcal{C}\to\mathcal{A}$ and $g\colon\mathcal{C}\to\mathcal{B}$ such
that, for all states $\rho$ and all $a\in\mathcal{A}$, $b\in\mathcal{B}$,
\begin{subequations}
\begin{align}
A_a(\rho) &= \sum_{c\in f^{-1}(a)} C_c(\rho),\\
B_b(\rho) &= \sum_{c\in g^{-1}(b)} C_c(\rho).
\end{align}
\end{subequations}
Thus, measuring $C$ and then applying $f$ and $g$ to the outcome
$c\in\mathcal{C}$ yields an outcome for $A$ and an outcome for $B$,
respectively, in such a way that the marginal statistics of $A$ (respectively,
$B$) are indistinguishable from those obtained by measuring $A$ (respectively,
$B$) alone.

The joint observable $J_{(a,b)}$ is related to the common refinement $C_c$ by
\begin{equation}
J_{(a,b)} = \sum_{c\in f^{-1}(a)\cap g^{-1}(b)} C_c.
\end{equation}
\end{definition}

The term refinement reflects the fact that $C$ is a finer observable
than both $A$ and $B$. Conversely, $A$ and $B$ are called
coarse-grainings of $C$, since they are obtained by grouping
(coarse-graining) the outcomes of $C$.

Therefore, $A$ and $B$ are jointly measurable if and only if they have a common
refinement. If $A$ and $B$ are not jointly measurable, they are said to be
incompatible.

\begin{definition}[Sequential measurements]
A sequential measurement of observables $A$ and $B$, with $A$ measured first
and $B$ second, is a procedure in which one first performs a measurement of $A$
on a physical system and then performs a measurement of $B$ on the same system,
so that the second measurement acts on the state left behind by the first.
\end{definition}

\begin{definition}[Non-disturbance]
Let $M$ be a measurement of observable $A$ and $N$ a measurement of observable
$B$. We say that $M$ does not disturb $N$ if, for every initial state $\rho$,
the outcome statistics of $N$ performed after $M$ are identical to those of $N$
performed on $\rho$ when no measurement of $A$ is performed. Otherwise, $M$ is
said to disturb $N$. We say that $M$ does not disturb observable $B$ if it does
not disturb any measurement $N$ of $B$.
\end{definition}

\begin{theorem}
Let $A$ and $B$ be observables. If $A$ and $B$ are incompatible, then for every measurement of $A$ there exists an initial state on which that measurement disturbs $B$.
\end{theorem}

\begin{proof}
Fix a measurement $M$ of $A$ and suppose that $M$ does not disturb $B$ for any initial state. Then, for every initial state, the outcome statistics of a subsequent measurement of $B$ are the same as if $M$ had not been performed. Consequently, the sequential procedure ``perform $M$, then measure $B$'' yields joint outcome statistics whose marginal for $A$ is the statistics of $A$ and whose marginal for $B$ is the statistics of $B$. Such a procedure therefore constitutes a common refinement of $A$ and $B$. But by definition of incompatibility, no common refinement of $A$ and $B$ exists. Hence $M$ must disturb $B$ on at least one initial state.
\end{proof}

Notice, however, that joint measurability of two observables does not imply that they can be measured sequentially, one after the other, without the first measurement disturbing the second.




%


\end{document}